\documentclass[12pt]{spieman}  
\usepackage{amsmath,amsfonts,amssymb}
\usepackage{graphicx}
\usepackage{setspace}
\usepackage{tocloft}
\usepackage{verbatim}
\usepackage{siunitx}
\usepackage[mathscr]{euscript}
\usepackage{mathtools}
\usepackage{bm}

\usepackage[dvipsnames]{xcolor}

\title{Characterizing deformable mirrors for the MagAO-X instrument}

\author[a,b,c,*]{Kyle Van Gorkom}
\author[a]{Jared R.\ Males}
\author[a]{Laird M.\ Close}
\author[a,b]{Jennifer Lumbres}
\author[a,b]{Alex Hedglen}
\author[a]{Joseph D.\ Long}
\author[a]{Sebastiaan Y.\ Haffert}
\author[a,b,d,e]{Olivier Guyon}
\author[a,b]{Maggie Kautz}
\author[a,b]{Lauren Schatz}
\author[f]{Kelsey Miller}
\author[a,b]{Alexander T. Rodack}
\author[a,b]{Justin M.\ Knight}
\author[a]{Katie M.\ Morzinski}
\affil[a]{University of Arizona, Steward Observatory, Tucson, Arizona, United States}
\affil[b]{University of Arizona, College of Optical Sciences, Tucson, Arizona, United States}
\affil[c]{NASA Goddard Space Flight Center, Greenbelt, Maryland, United States}
\affil[d]{National Institutes of Natural Sciences, Subaru Telescope, National Observatory of Japan, Hilo, Hawaii, United States}
\affil[e]{National Institutes of Natural Sciences, Astrobiology Center, Mitaka, Japan}
\affil[f]{Air Force Research Laboratory, Albuquerque, New Mexico, United States}

\usepackage{lineno}

\let\oldequation\align
\let\oldendequation\endalign

\renewenvironment{align}
  {\linenomathNonumbers\oldequation}
  {\oldendequation\endlinenomath}

\newcommand*\dif{\mathop{}\!\mathrm{d}}

\newcommand{\rev}{\ignorespaces}

\cftpagenumbersoff{figure}
\cftpagenumbersoff{table} 
\begin{document} 
\maketitle

\begin{abstract}
The MagAO-X instrument is a new extreme adaptive optics system for high-contrast imaging at visible and near-infrared wavelengths on the Magellan Clay Telescope. A central component of this system is a 2040-actuator microelectromechanical deformable mirror (DM) from Boston Micromachines Corp.\ that operates at 3.63 kHz for high-order wavefront control (the tweeter). Two additional DMs from ALPAO perform the low-order (the woofer) and non-common-path science-arm wavefront correction (the NCPC DM). Prior to integration with the instrument, we characterized these devices using a Zygo Verifire Interferometer to measure each DM surface. We present the results of the characterization effort here, demonstrating the ability to drive tweeter to a flat of 6.9 nm root mean square (RMS) surface (and 0.56 nm RMS surface within its control bandwidth), the woofer to 2.2 nm RMS surface, and the NCPC DM to 2.1 nm RMS surface over the MagAO-X beam footprint on each device. Using focus-diversity phase retrieval on the MagAO-X science cameras to estimate the internal instrument wavefront error (WFE), we further show that the integrated DMs correct the instrument WFE to 18.7 nm RMS, which, combined with a 11.7\% pupil amplitude RMS, produces a Strehl ratio of 0.94 at H$\alpha$. 
\end{abstract}


{\noindent \footnotesize\textbf{*}Kyle Van Gorkom,  \linkable{kyle.vangorkom@nasa.gov} }

\begin{spacing}{2}   

\section{Introduction}
\label{sect:intro}  

\begin{figure}
\begin{center}
\begin{tabular}{c}
\includegraphics[width=6.75in]{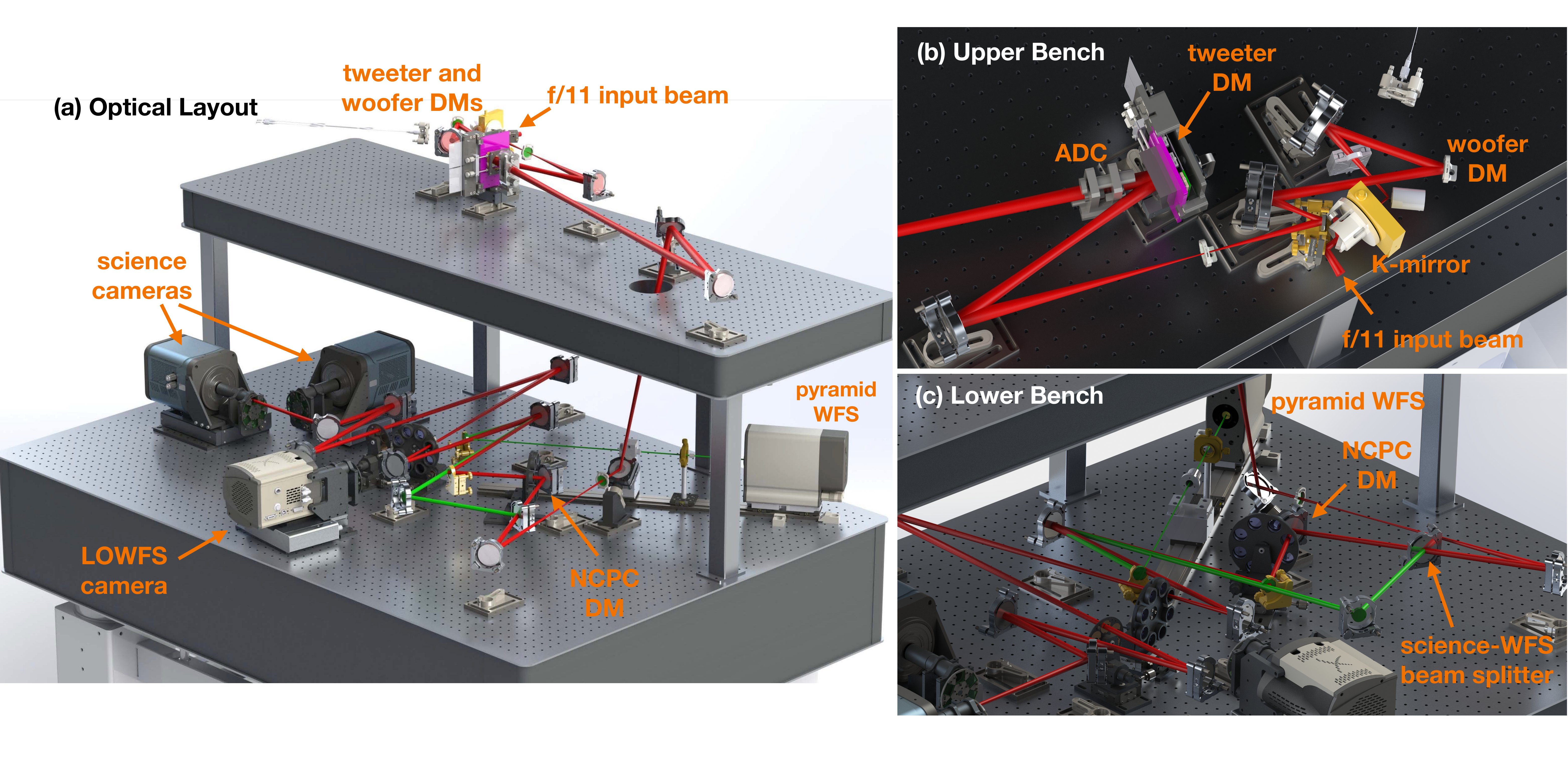}
\end{tabular}
\end{center}
\caption 
{ \label{fig:dm_images}
\rev{(a) Render of the MagAO-X instrument, with close-ups of the (b) upper and (c) lower benches. The tweeter and woofer DMs are located on the upper bench, and the NCPC DM is located on the lower bench after the beam splitter between the science and WFS paths.}} 
\end{figure}

MagAO-X\cite{malesspie,lairdspie} is an extreme adaptive optics (ExAO) instrument for the 6.5m Magellan Clay telescope at the Las Campanas Observatory in Chile designed for high contrast imaging (HCI) at visible to near-infrared wavelengths. Once instrument commissioning is complete, MagAO-X will deliver a Strehl ratio of $\gtrsim 0.7$ at H$\alpha$ (0.656$\si{\micro\meter}$), 14-30 mas resolution, and $10^{-4}$ contrast from $\sim$1 to 10 $\lambda$/D. In its current configuration, a vector Apodizing Phase Plate (vAPP)\cite{snik} coronagraph performs the starlight suppression for HCI, while a future upgrade will introduce a phase induced amplitude apodization complex-mask coronagraph (\nolinebreak{PIAACMC})\cite{Guyon2010}. The main adaptive optics (AO) loop comprises two deformable mirrors (DMs) in an offloading woofer-tweeter control scheme\cite{brennan}. A third DM downstream of the AO loop is dedicated to non-common-path correction (NCPC). The system woofer, a high-stroke 11x11 97-actuator deformable mirror from ALPAO SAS, provides low-order wavefront correction, while the tweeter---a 50x50 2040-actuator (2K) microelectromechanical systems (MEMS) DM from Boston Micromachines Corp.\ (BMC)---simultaneously provides high-order wavefront correction. A pyramid wavefront sensor (PyWFS)\cite{schatz} operated at up to 3.6 kHz drives these DMs. The downstream DM, a second ALPAO DM97, performs the NCP correction with low-order and focal-plane wavefront-sensing in the coronagraph arm\cite{miller, Miller2019}. See Figure \ref{fig:dm_images}.

High-order MEMS DMs from BMC are currently in use on other ExAO instruments, including GPI\cite{poyneer_mems} (Gemini Planet Imager) and SCExAO\cite{scexao} (Subaru Coronagraphic Extreme Adaptive Optics instrument). The characterization and on-sky performance of these devices has been extensively reported\cite{Morzinski2006,Evans2006,norton2009,Blain2010}. ALPAO devices have seen adoption in AO instruments including Raven\cite{Lamb2015} at the Subaru Telescope and NAOMI\cite{LeBouquin2018} (New Adaptive Optics Module for Interferometry) at the VLTI (Very Large Telescope Interferometer), and extensive characterization efforts of these devices have been likewise undertaken\cite{Vidal2019}.

\rev{MagAO-X represents the first application of ALPAO devices in an ExAO system. Here we report the characterization of recent BMC and ALPAO DM offerings for use in HCI, as well as a set of procedures followed to characterize and optimize these devices for ExAO both prior to and after integration with the instrument.} Section \ref{sect:char} describes the characterization testbed and procedures performed at the Arizona Extreme Wavefront Control Lab (XWCL)\cite{mespie}. Sections \ref{sect:tweeter} and \ref{sec:woofer_ncpc} report the results of characterization for the BMC and ALPAO DMs, and in Section \ref{sect:instrument} we discuss DM performance in the instrument and the optimization of internal instrument wavefront error.

\section{Characterization testbed and procedures}
\label{sect:char}

\begin{figure}
\begin{center}
\begin{tabular}{c}
\includegraphics[width=6.75 in]{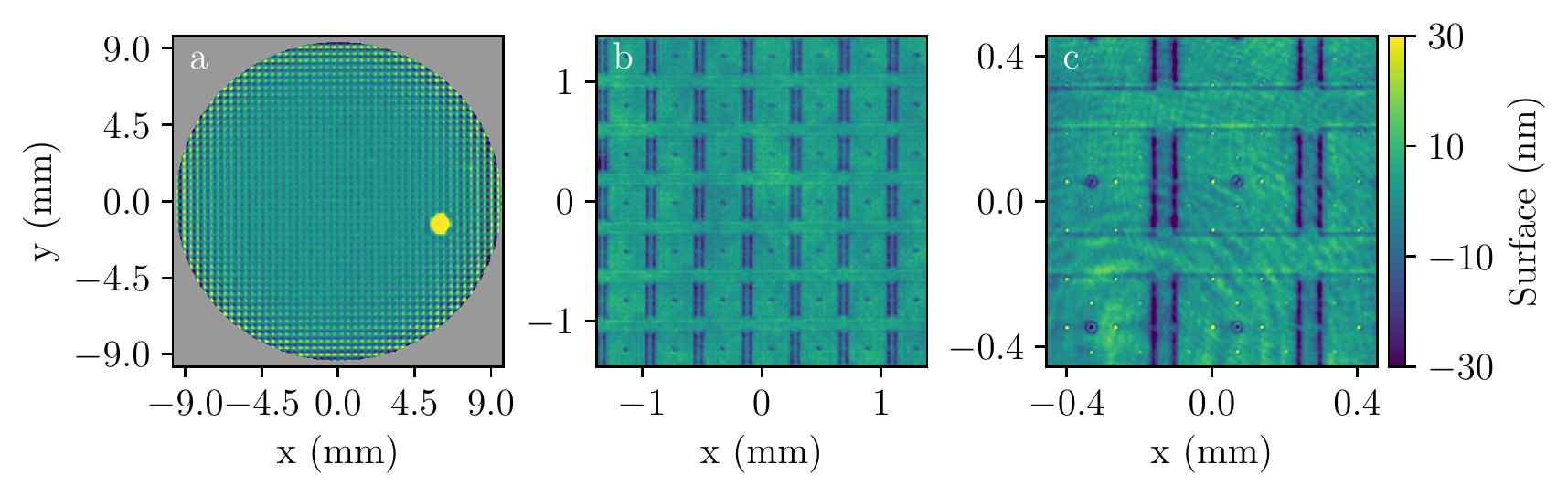}
\end{tabular}
\end{center}
\caption 
{ \label{fig:config_zooms}
Zygo surface measurements of the BMC 2K at (a) 31 $\si{\micro\meter}$/pixel, (b) 2.7 $\si{\micro\meter}$/pixel, and (c) 0.89 $\si{\micro\meter}$/pixel resolutions. Configurations (b) and (c) make use of two-lens beam reducers to achieve additional magnification. Low-order aberrations introduced by the lenses have been removed in post-processing.} 
\end{figure}

The characterization testbed employed a Zygo Verifire, a Fizeau-type interferometer with a HeNe laser at 632.8 nm, to make DM surface measurements. To reduce static aberrations in the interferometric reference, we used a 4 inch, $\lambda/50$ (PV) Dynaflect transmission flat. The internal optics of the Verifire enable up to 6x zoom, but some characterization activities required additional resolution to resolve closely-spaced fringes created by large strokes. To increase the resolution, we placed two-lens beam reducers in the collimated beam between the interferometer and DM, which resulted in an additional 6x and 20x increase in resolution. \rev{The beam reducers introduced low-order aberrations, but the characterization activities which made use of them (primarily localized measurements of inter-actuator and single-actuator stroke) were differential in nature, and the aberrations were removed by subtraction of the uncommanded surface or else by fitting Zernike polynomials in post-processing.} Example measurements demonstrating the magnification in the different configurations are shown in Fig.~\ref{fig:config_zooms}. 

To expedite the measurement process, we wrote a Python library\cite{zygo_code} to automate the interferometer measurements and DM commanding and synchronize the processes across multiple servers via a shared network drive. Zygo provides a Python API for interferometer measurement acquisition, basic post-processing, and writing to disk. Each DM has a vendor-provided API, which we interfaced with the Compute and Control for Adaptive Optics (\texttt{cacao}) package for low-latency DM control~\cite{guyon}.

\begin{figure}
\begin{center}
\includegraphics[width=6.75 in]{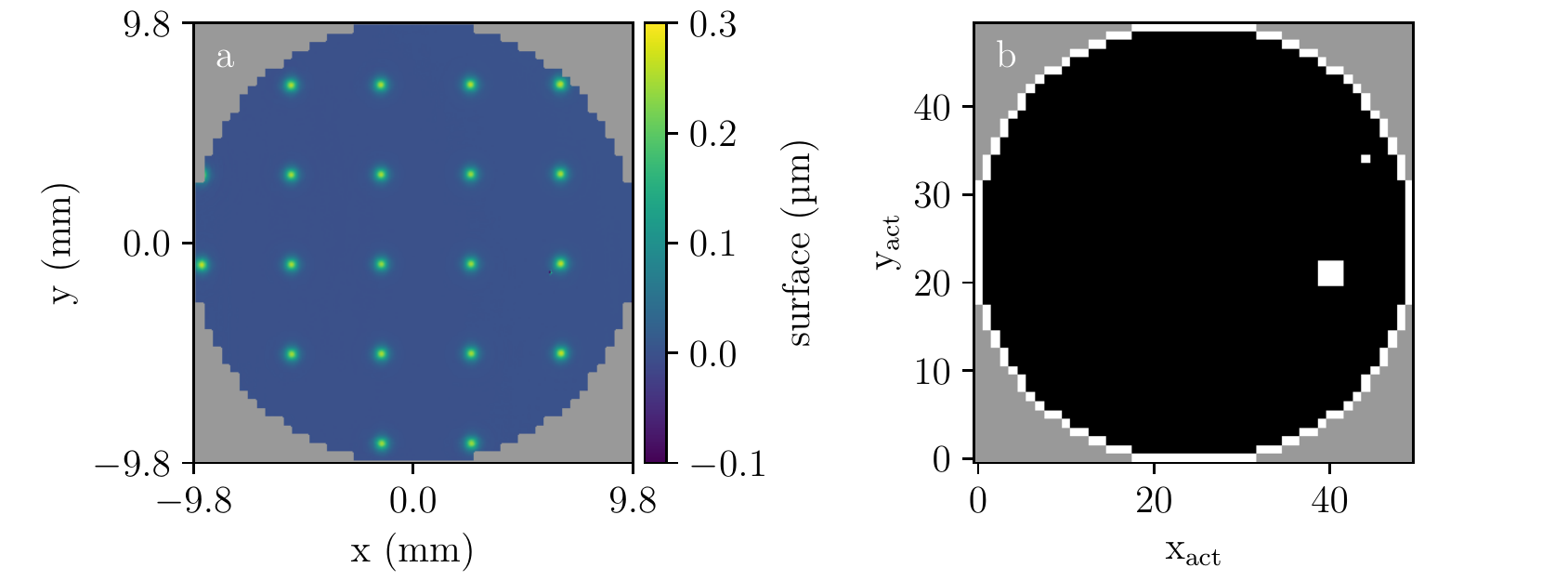}
 \end{center}
\caption 
{ \label{fig:ifgrid_slaved}
(a) An example interferometric measurement of a grid with every 10th actuator poked on the 2K DM, \rev{collected as part of the procedure to build an interaction matrix}. (b) The map of couple-controlled actuators (in white) adopted for closed-loop operations. \rev{All edge actuators are couple-controlled, along with a 3x3 region around the bump and an isolated actuator physically coupled to its neighbor.}} 
\end{figure}

In the case of the 97-actuator ALPAOs, we measured the influence functions (IFs, the characteristic response of the surface to a single-actuator command) individually. To accelerate the process for the 2040-actuator BMC, we measured the IFs in a grid pattern, with sufficient separation between commanded actuators to minimize coupling effects, and later extracted the individual IFs from the measurements of the grid pattern (see Figure \ref{fig:ifgrid_slaved}). \rev{We conservatively chose a grid with a pitch of 10 actuators to avoid concerns regarding coupling between actuators; however, additional time savings could be found with a more detailed study to identify the densest grid pattern that still maintains negligible coupling between poked actuators.} In each case, we measured positive and negative commands and computed the difference to remove the contribution of the static DM surface. Residual tip/tilt errors from mechanical disturbances on the testbed and \rev{linear} drift over time were fit and removed in post processing.

The resulting IF cubes were flattened into interaction matrices and (pseudo-)inverted to create command matrices for closed-loop characterization activities. \rev{In this measurement configuration, all actuators are sampled by the Zygo interferometer and no single influence function can be constructed from a linear combination of influence functions from other actuators, so we expect the interaction matrices to be full rank.} To minimize edge effects from actuators on the perimeter of the DMs, we thresholded on the RMS of the IF measurements to create a map of couple-controlled actuators that were not commanded directly but instead assigned commands from a weighted combination of low-order Zernikes fit to the command map and the mean of nearest-neighbors (Figure \ref{fig:ifgrid_slaved}). Flat maps later used during alignment and integration with MagAO-X were created in closed loop in front of the interferometer.

The 2K DM must be operated at relative humidity levels $< 30\%$ to avoid actuator oxidation damage~\cite{morzinski_2012}. \rev{To meet this requirement, we ran nitrogen through the windowed enclosure that seals the DM and measured the humidity via a sensor on the outlet port.} The ALPAO devices have no equivalent requirement on relative humidity. Additional details describing the the particular measurements and procedures required to characterize each DM are described in the following sections.

\section{Tweeter characterization}
\label{sect:tweeter}

\begin{figure}
\begin{center}
\includegraphics{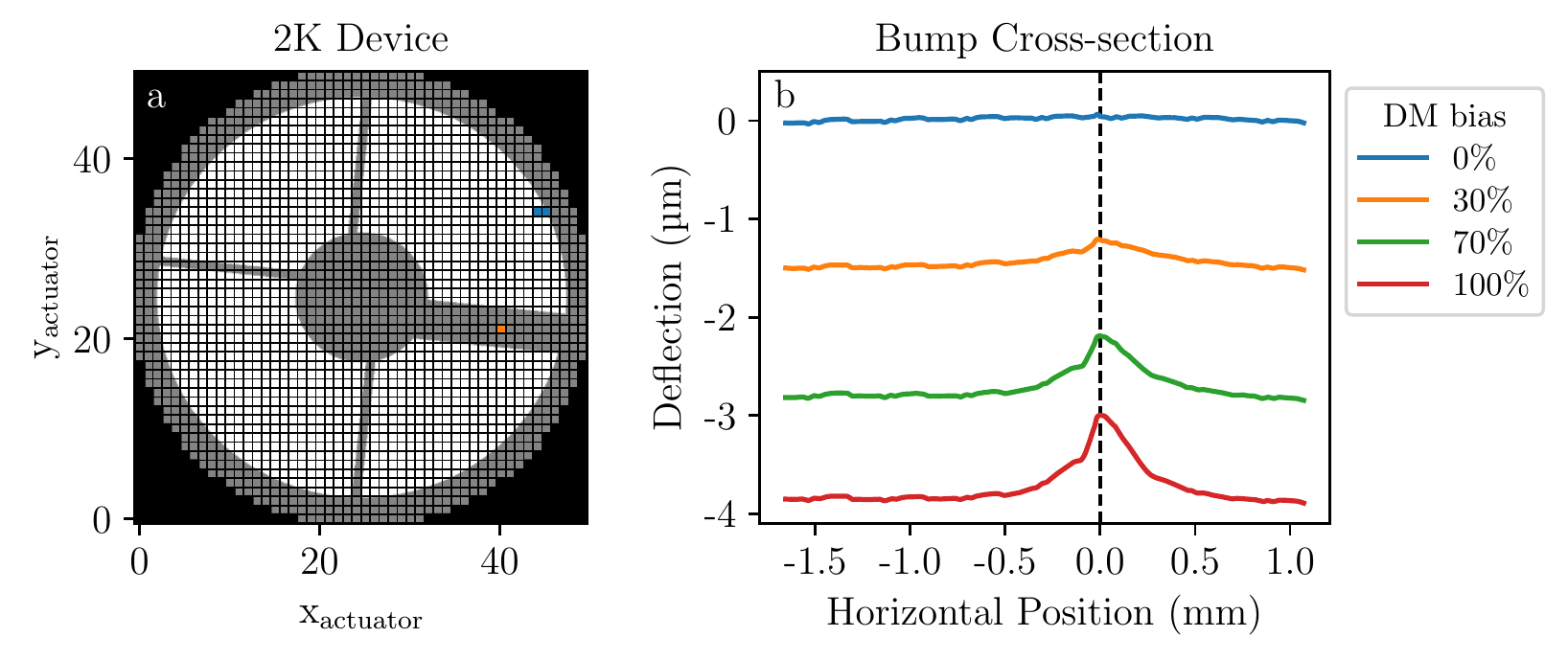}
\end{center}
\caption 
{ \label{fig:2k_device}
(a) The geometry of the 2K DM actuator grid. The projected vAPP pupil\rev{, rotated to place the bump behind the large spider arm,} is overlaid in grey. The two coupled actuators are indicated in blue, and the location of the bump is indicated in orange. (b) Cross-section of the 2K surface through the bump. As the DM surface is biased and deflected downwards, the reflective membrane at the location of the bump lags behind the surrounding surface.} 
\end{figure}

The system tweeter is a 2040-actuator 3.5 $\si{\micro\meter}$ stroke MEMS DM from Boston Micromachines Corporation. The actuators are arranged on a square grid with a 400 $\si{\micro\meter}$ pitch inscribed in a 19.6 mm diameter aperture. The reflective membrane has an unprotected gold coating, and the device is environmentally sealed behind an AR-coated window with a $6^\circ$ wedge. \rev{Even though MagAO-X is designed for visible wavelengths, the unprotected gold coating provides ${>}95$\% reflectance at H$\alpha$, ${>}50\%$ at $0.5 \si{\micro\meter}$, and ${\sim} 40$\% down to $0.4\si{\micro\meter}$, yielding excellent quantum efficiency over the bandpasses of scientific interest}. The maximum safe voltage is capped at 210 V. This is a 100\% yield device (no stuck or nonresponsive actuators) with a few notable defects: two neighboring actuators are coupled (both respond when a voltage is applied to either one), and a bump on the surface appears and grows as the DM surface is deflected downwards, to a maximum height of 0.86 $\si{\micro\meter}$ above the surrounding surface and a width of 1.4 mm. These features are shown in Figure \ref{fig:2k_device}, along with the coronagraphic pupil, which is rotated to block the light scattered from the bump.

\subsection{Tweeter stroke metrics}

\begin{figure}
\begin{center}
\includegraphics[width=6.75in]{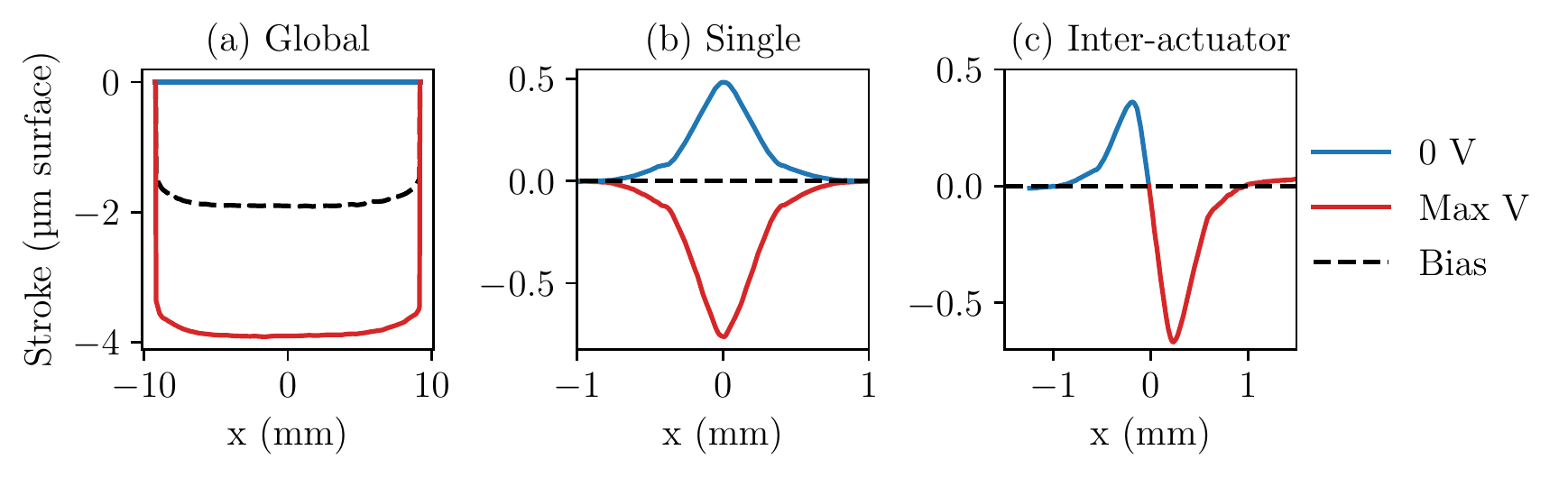}
\end{center}
\caption 
{ \label{fig:stroke_description}
Cross-sections of the tweeter DM surface to demonstrate the stroke definitions adopted in this work. Dashed black lines show the surface position at a given voltage bias, blue lines show the surface in a released (0 voltage) state, and red lines show the surface in a railed (maximum voltage) state. For global stroke (a), all actuators are commanded together; for single-actuator stroke (b), a single actuator is released and then railed; for inter-actuator stroke (c), two neighboring actuators are commanded in opposite directions. The stroke in each case is defined to be twice the smaller of the two surface deflections from the bias position (Equation \ref{eqn:stroke_def}).} 
\end{figure}

In an HCI system, actuator saturation scatters light into the dark hole region and can significantly reduce contrast\cite{morzinski2008}. The woofer-tweeter architecture reduces the stroke requirement on the 2K, and the choice of 2K operating bias has significant implications for the available stroke and probability of saturation. The stroke of a MEMS device depends heavily on the overall voltage bias, \rev{the voltage applied across the entire DM to move the surface to an intermediate position and allow both positive and negative deflections}. We investigate three stroke metrics (reported in this work in terms of surface rather than wavefront) in order to evaluate 2K performance and adopt an operational bias. The first metric---single-actuator stroke---describes the full range of surface deflection from a bias position when a single actuator is released (0 V) and then subsequently railed (210 V). The deflection in the two directions is highly asymmetric for most bias positions (and entirely one-sided at 0 and 210 V); since the ability to move in both directions is operationally important, we capture the effective dynamic range of the deflection by taking the stroke $s$ to be twice the smaller of the two surface deflections:
\begin{align}\label{eqn:stroke_def}
s = 2 \times \min( |s_\mathrm{railed}|,\ |s_\mathrm{released}|)
\end{align}

Inter-actuator stroke measures the surface peak-to-valley (PV) of two neighboring actuators when one is railed and the other released, and global stroke captures the range of surface deflection when all actuators are released and railed together. The same symmetry constraint described above is applied to these metrics. These stroke definitions are shown in Fig.~\ref{fig:stroke_description}.

\begin{figure}
\begin{center}
\includegraphics{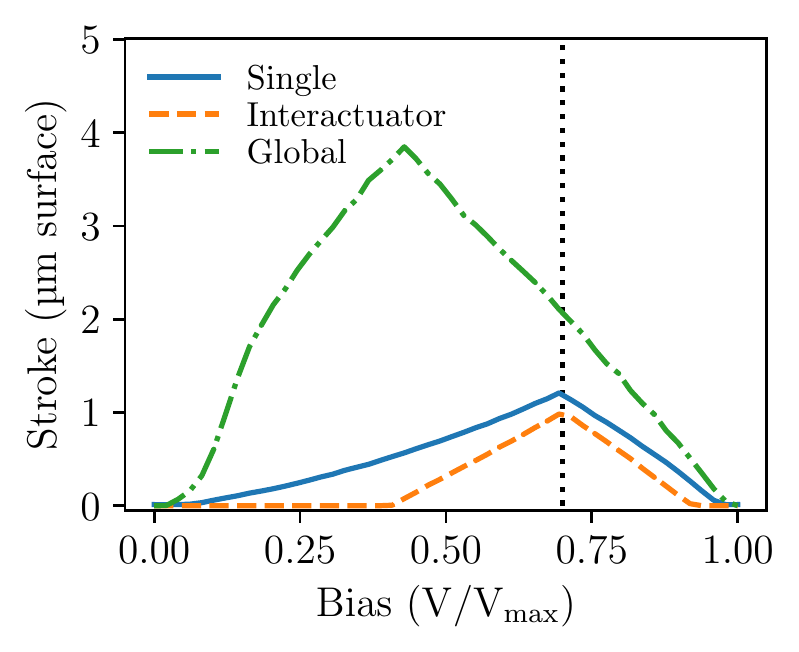}
\end{center}
\caption 
{ \label{fig:stroke_metrics}
Three metrics of tweeter stroke as a function of bias. Single- and inter-actuator stroke are maximized at 70\% voltage bias (dashed vertical line), while global stroke is maximized at 40\% bias.} 
\end{figure}

In a testbed configuration with 2.7$\si{\micro\meter}$/pix resolution, we measured single-actuator and inter-actuator stroke at 8 locations distributed across the DM surface and found similar behaviors for all. The median of each of these curves is plotted in Figure \ref{fig:stroke_metrics}. Global stroke peaks at $\sim 4 \si{\micro\meter}$ surface at $40\%$ bias, while single- and inter-actuator stroke peak at 1.2 and 1 $\si{\micro\meter}$ surface, respectively, at $70\%$ bias. Inter-actuator stroke places an upper limit on the amplitude of the highest spatial frequencies that can be corrected with the tweeter and drives our choice of operational bias to be 70\%.

\begin{figure}
\begin{center}
\includegraphics{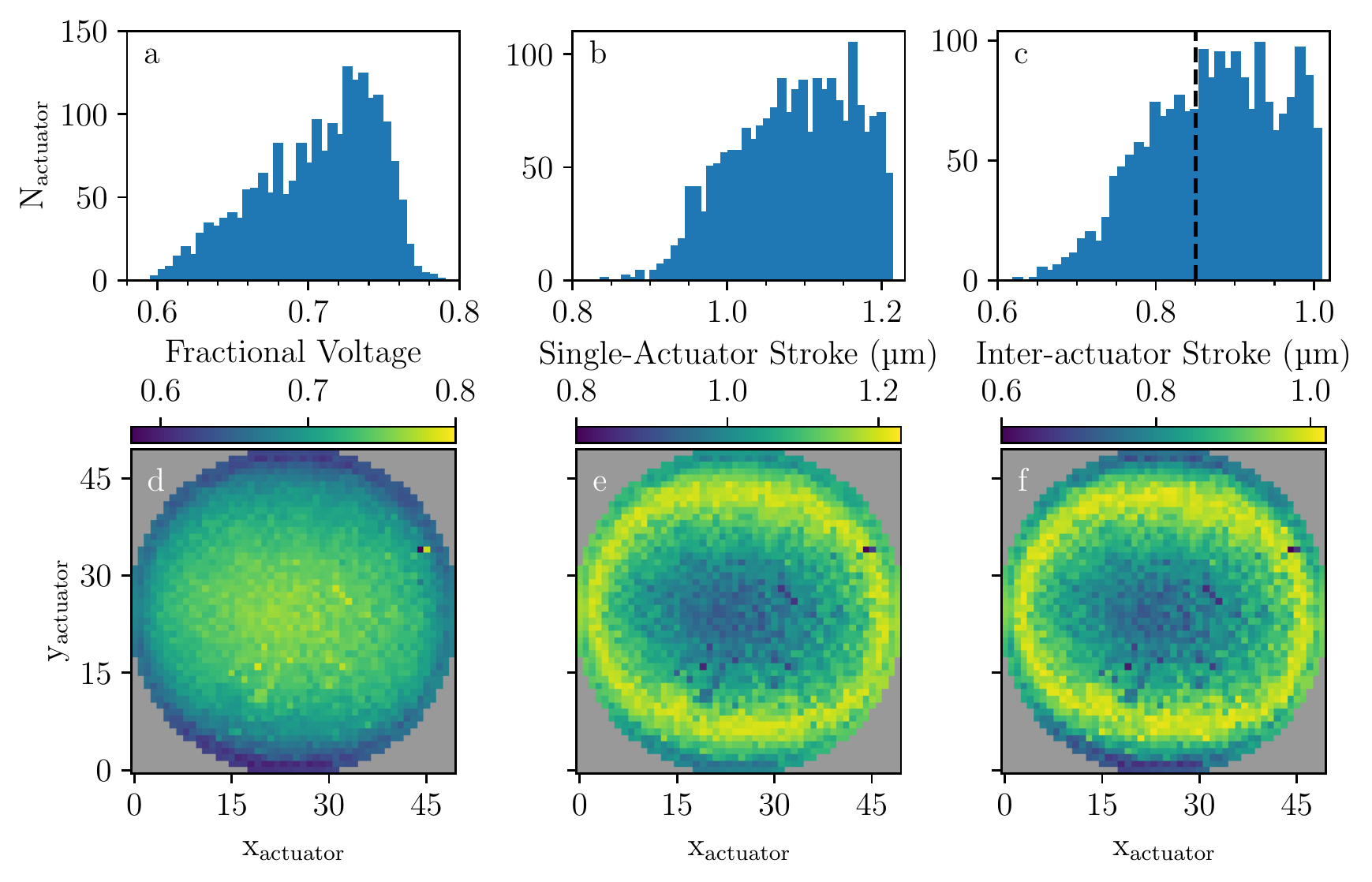}
\end{center}
\caption 
{ \label{fig:fullflat_metrics}
Tweeter (BMC 2K) voltage and stroke metrics for an absolute flat defined at 70\% bias. Top: Histograms showing the distribution of actuators by (a) fractional voltage required to achieve this flat, (b) available single-actuator stroke, and (c) available inter-actuator stroke. Bottom: Corresponding maps of the spatial distributions of (d) fractional voltage, (e) single-actuator stroke, and (f) inter-actuator stroke. The dashed line shows the minimum required inter-actuator stroke\cite{males}. With this flat definition, only $63\%$ of actuators meet this requirement. Compare to Figure \ref{fig:relaxedflat_metrics}.} 
\end{figure}

\begin{figure}
\begin{center}
\includegraphics{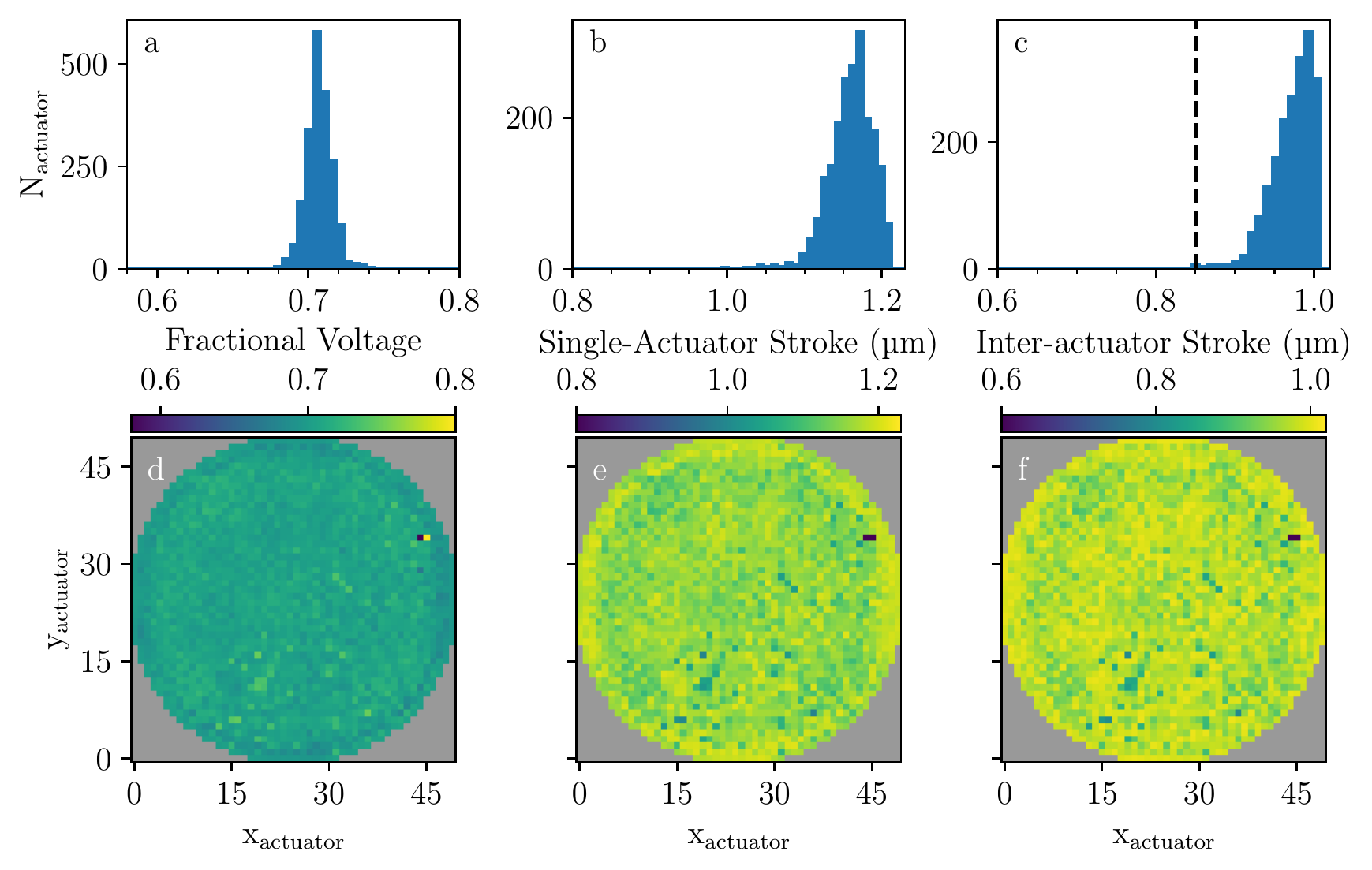}
\end{center}
\caption 
{ \label{fig:relaxedflat_metrics}
Tweeter voltage and stroke metrics for a ``relaxed'' flat defined at 70\% bias, where low-orders have been left uncorrected to be offloaded to the woofer. Top: Histograms showing the distribution of actuators by (a) fractional voltage required to achieve this flat, (b) available single-actuator stroke, and (c) available inter-actuator stroke. Bottom: Corresponding maps of the spatial distributions of (d) fractional voltage, (e) single-actuator stroke, and (f) inter-actuator stroke. The dashed line shows the minimum required inter-actuator stroke\cite{males}. With this flat definition, $>99\%$ of actuators meet this requirement.  Compare to Figure \ref{fig:fullflat_metrics}.} 
\end{figure}

When all actuators are commanded to their 70\% bias position, the tweeter DM has 120 nm RMS surface of low-order sag. Driving the DM to an absolute flat state from this bias requires a large range of voltages (from 60 to 80\%, or 126-168V). As a result, a large fraction of the actuators are at a non-optimal bias for single- and inter-actuator stroke (see Figure \ref{fig:fullflat_metrics}). In this configuration, only $63\%$ of the actuators are estimated to have inter-actuator stroke $> 0.85 \si{\micro\meter}$ surface. If instead the static tweeter sag is offloaded to the woofer and only high-order static errors are corrected on the 2K, the range of voltages in the flat bias can be narrowed about the optimal point, driving single- and inter-actuator stroke up across the DM surface. \rev{This ``relaxed'' flat leaves 103 nm RMS surface of uncorrected sag on the tweeter, of which 102.6 nm RMS is within the woofer control bandwidth and well within the available stroke for low-order modes on the woofer (see Section \ref{sec:woofer_ncpc})}. In this configuration, $>99\%$ of the actuators exceed the $0.85 \si{\micro\meter}$ surface requirement\cite{males} on inter-actuator stroke (see Figure \ref{fig:relaxedflat_metrics}). We adopted this latter flat when integrating the 2K into the instrument to maximize the available tweeter stroke.

\subsection{Tweeter surface quality}\label{sec:tweeter_surface_quality}

\begin{figure}
\begin{center}
\includegraphics{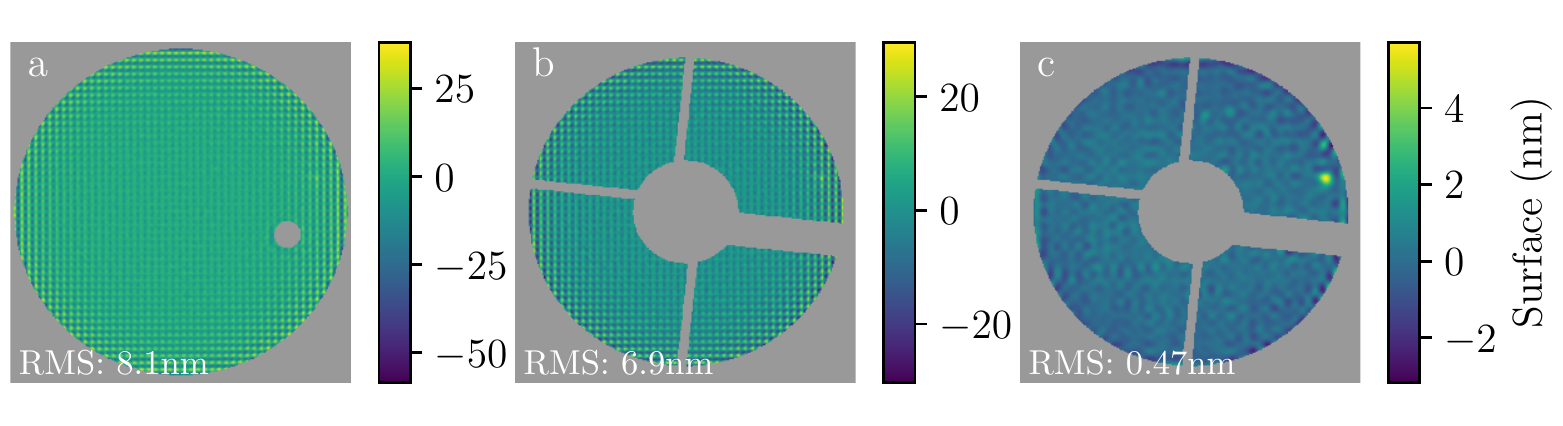}
\end{center}
\caption 
{ \label{fig:tweeter_surface}
Tweeter (BMC 2K): Closed-loop flats measured and defined on the Zygo interferometer. (a) The full DM flat with an RMS of 8.1 nm surface (12.6 nm RMS surface with the bump). (b) The flat within the projected vAPP pupil, with an RMS of 6.9 nm surface. (c) The in-band surface within the projected vAPP pupil produced via Butterworth filter with a 24 cycle/aperture cut-off, with an RMS of 0.47 nm surface.} 
\end{figure}

The closed-loop tweeter flat defined in front of the Zygo is shown in Figure \ref{fig:tweeter_surface}.  Over the active aperture of the DM, the flat has 12.6 nm RMS surface or 8.1 nm RMS surface with the bump masked \rev{in post-processing}.  Over the MagAO-X vAPP pupil \rev{rotated to obscure the bump}, the RMS reduces to 6.9 nm RMS surface. The in-band RMS of the flat is estimated in two ways. In the first, the in-band RMS is computed from the two-dimensional power spectral density (PSD) following
\begin{align}\label{eqn:psd}
\mathrm{\sigma} = \sqrt{ \frac{1} {N_x N_y} \sum_{\vec{k}_i}^{\vec{k}_\mathrm{cutoff}} \mathrm{PSD}(\vec{k}_i) \Delta k_y \Delta k_y} 	
\end{align}
where $N_x$ and $N_y$ are the number of pixels in $x$ and $y$, $\vec{k}_i = (k_x, k_y)$ is the spatial frequency variable, $\Delta k_y$ and $\Delta k_x$ are the frequency-space samplings, and $\vec{k}_\mathrm{cutoff}$ is the maximum controllable frequency, set by the Nyquist sampling of the DM actuators. \rev{To suppress spurious frequency components from the hard cutoff at the edge of the aperture, we apply a window function prior to taking the fast Fourier transform (FFT). The actuator-scale print-through pattern on the tweeter, however, is stronger at larger radii (see Figure \ref{fig:tweeter_surface}a and \ref{fig:tweeter_surface}b), and window functions that taper off well before the edge of the aperture (for example, a radial Hann or Hamming window) tend to underestimate the print-through amplitude. By experimental exploration of the choice of window function and its effect on the trade-off between capturing the print-through pattern and suppressing frequencies from the aperture function, we adopted the radial Tukey window (cosine fraction parameter $\alpha=0.3$).}

In the second approach, the in-band surface is estimated by applying a two-dimensional, 5th-order Butterworth filter with the DM cutoff frequency in the Fourier domain before inverse-transforming to the spatial domain. \rev{The filter order was chosen heuristically to minimize the out-of-band surface content passed by the filter without introducing ringing in the in-band surface with a hard cutoff. The use of the Butterworth filter here is primarily intended to provide a visualization of the in-band surface and likely produces a slight overestimate of the in-band RMS.}

\begin{figure}
\begin{center}
\includegraphics{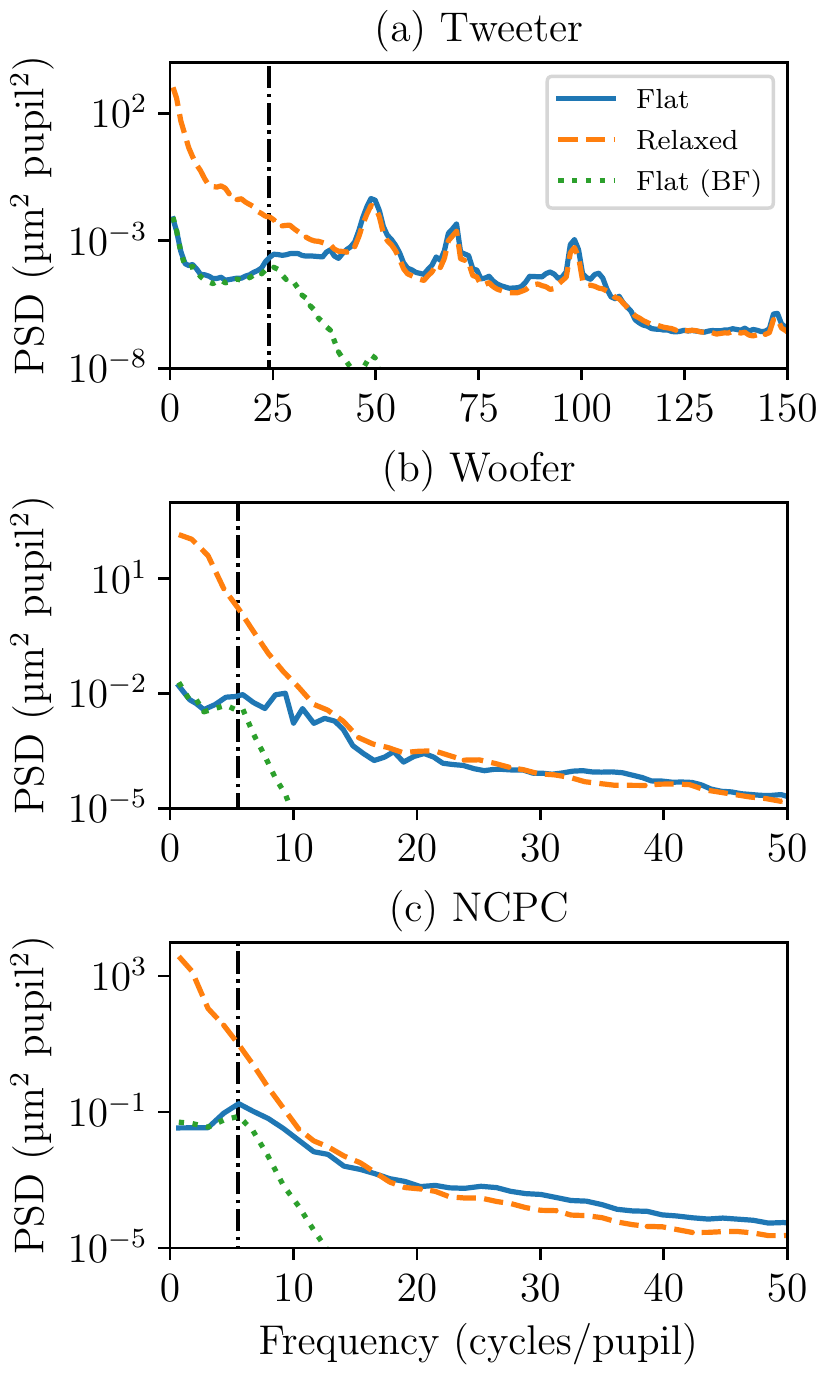}
\end{center}
\caption 
{ \label{fig:DMpsds}
PSDs for the (a) tweeter, (b) woofer, and (c) NCPC DMs, computed from Zygo flat surface measurements. The ``relaxed`` surface PSDs (orange dashed lines) are computed from measurements of each DM powered-on and at its operating bias position prior to flattening. The solid lines plot the PSDs computed from the best flats, and the flat PSDs after Butterworth filtering (BF) are included for comparison. The vertical dash-dot lines indicate the cutoff frequency for each DM.} 
\end{figure}

The tweeter in-band (24 cycles across the pupil) RMS is estimated from Equation \ref{eqn:psd} to be 0.56nm RMS surface, or 0.94 nm RMS surface via Butterworth filtering (and 0.47 nm RMS surface within the vAPP pupil). One-dimensional PSDs are calculated from the radial average of the two-dimensional PSD and are shown in Figure \ref{fig:DMpsds}, in both the relaxed state and powered with an active flat.

\section{Woofer and NCPC characterization}
\label{sec:woofer_ncpc}

\begin{figure}
\begin{center}
\includegraphics{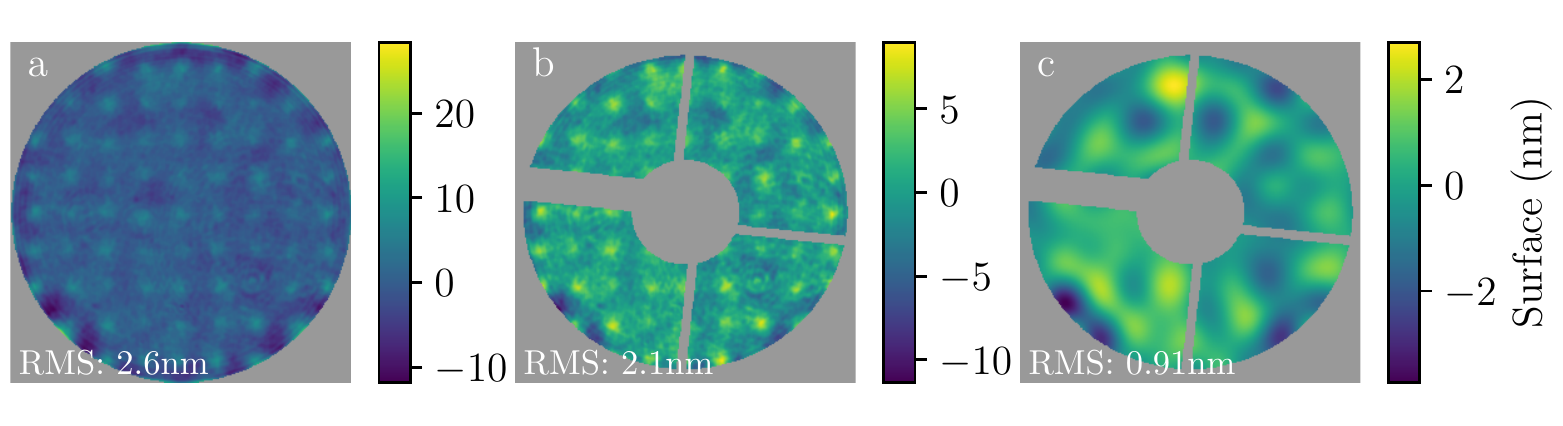}
\end{center}
\caption 
{ \label{fig:woofer_flats}
Woofer (ALPAO DM97): Closed-loop flats measured and defined on the Zygo interferometer. (a) The full DM surface with an RMS of 2.6 nm surface. (b) The surface within the projected vAPP pupil, with an RMS of 2.1 nm surface. (c) The in-band surface produced via Butterworth filter with a 5.5 cycle/aperture cut-off frequency, with an RMS of 0.91 nm surface.} 
\end{figure}

\begin{figure}
\begin{center}
\includegraphics{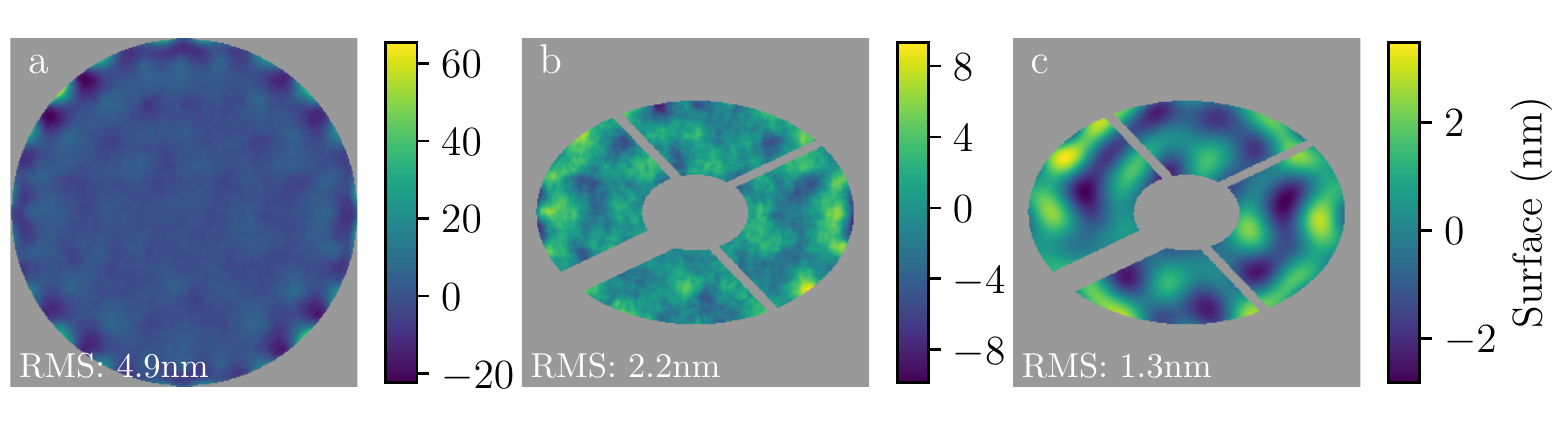}
\end{center}
\caption 
{ \label{fig:ncpc_flats}
NCPC (ALPAO DM97): Closed-loop flats measured and defined on the Zygo interferometer. (a) The full DM surface with an RMS of 4.9 nm surface. (b) The surface within the projected vAPP pupil, with an RMS of 2.2 nm surface. (c) The in-band surface produced via Butterworth filter with a 5.5 cycle/aperture cut-off frequency, with an RMS of 1.3 nm surface.} 
\end{figure}

The woofer and NCPC DMs are two ALPAO DM97 devices, each with 97 actuators arranged in an 11x11 grid.  The DMs are compact voice coil continuous facesheet deformable mirrors with a 13.5 mm aperture and actuators on a 1.5 mm pitch. The reflective membrane is protected silver. These are high-stroke devices, capable of $12-15 \si{\micro\meter}$ surface P-V of tip/tilt and $\ge 10 \si{\micro\meter}$ surface P-V of low-order Zernike modes. \rev{The NCPC has $1.52 \si{\micro\meter}$ inter-actuator stroke (surface), while the woofer has $2.24 \si{\micro\meter}$ and was selected as the system woofer for this reason.} Both devices can be run at $>$2 kHz. We flattened both DMs following the procedure described in Section ~\ref{sect:char}; in this case, the surface at the edge of the pupil was well-behaved enough that the closed loop didn't require couple-controlling any edge actuators, although some edge ringing can be observed in the flats. Over the full DM aperture, the woofer achieved of flat of 2.6 nm RMS, while the NCPC DM achieved a slightly higher error of 4.9 nm RMS. \rev{The small difference in the flat quality is likely due to unspecified variations in the product delivered by ALPAO; both devices, however, meet the surface quality specifications provided to the manufacturer.} Over the vAPP pupil projected onto the DM surfaces, both flats reduce to $\sim 2$nm RMS. Within the control radius of the ALPAO DMs, the flat error over the full aperture is 1.4 and 3.5 nm RMS for the woofer and NCPC, respectively (computed by Equation \ref{eqn:psd}), or 0.91 and 1.25 nm RMS as computed via the Butterworth filtering procedure. PSDs for both devices are reported in Figure \ref{fig:DMpsds}. See Figures \ref{fig:woofer_flats} and \ref{fig:ncpc_flats} for ALPAO surface maps.

\subsection{ALPAO shape creep}

\begin{figure}
\begin{center}
\includegraphics{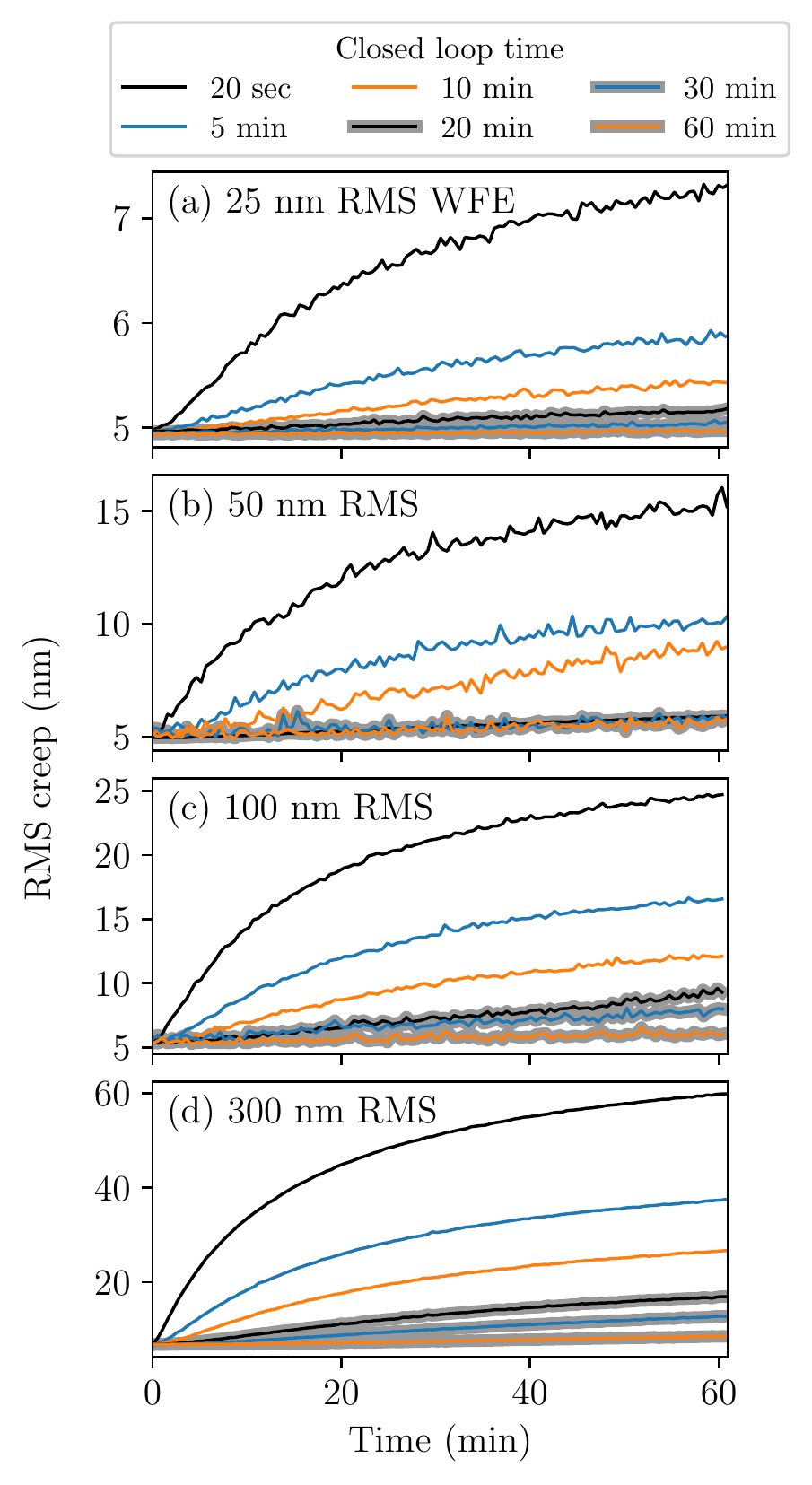}
\end{center}
\caption 
{ \label{fig:creep}
ALPAO shape creep after opening the loop on (a) 25 nm, (b) 50 nm, (c) 100 nm, and (d) 300 nm RMS low-order wavefront errors on the DM. Each line shows the mean surface creep of 5 realizations of WFE after closing the loop on the shape for the indicated time period. Longer closed-loop periods mitigate the observed creep once the loop is opened \cite{ALPAO2017}. The 5 nm RMS floor arises from surface errors outside of the control bandwidth.} 
\end{figure}

\rev{The surface shape of many ALPAO DMs have been noted to slowly drift away from a commanded shape in a manner influenced by the history of shapes previously held on the DM surface, an effect known as creep\cite{Bitenc2014,Bitenc2017,Lamb2015}, which ALPAO attributes to the slow relaxation of the polymer material of the underlying actuator plate\cite{ALPAO2017}.} Bitenc et al.\cite{Bitenc2014,Bitenc2017} developed a software compensation scheme for creep that takes into account the history of commands placed on the DM and applies a time-dependent correction factor, and also identified an independent source of surface error credited to the heating of the electrical components that drive the DM. Actuator and modal gains are also known to depend on device temperature and can be compensated for with calibrated look-up tables\cite{LeBouquin2018, ALPAO2017}, although on-sky tests have shown that modal gains are uncorrelated with the DM housing temperature and are better estimated on-sky in the absence of an internal temperature sensor\cite{Woillez2019}.

The \rev{drift} of the DM membrane via creep contributes a time-dependent source of WFE to the system. In closed loop, creep is corrected as it arises, so we do not anticipate the need for a creep compensation scheme on the system woofer or the NCPC DM under most conditions. In some observing modes, however, we may want to maintain the NCPC DM in an optimized shape that minimizes NCP aberrations without sacrificing science photons to maintain an active NCP loop. In this scenario, maintaining a creep-free DM shape is desirable. By closing the loop on the NCP WFE for a time period before fixing the DM on the optimized shape, the effect of creep-induced WFE can be reduced. Fig.~\ref{fig:creep} shows the RMS WFE creep after opening the loop, for different choices of closed-loop time periods and for different levels of initial WFE. Each curve is the mean creep observed for five realizations of low-order (the first 10 Zernikes, neglecting piston) WFE. Longer closed-loop periods result in a more stable surface at the cost of observing efficiency. At the expected 25 nm RMS NCP WFE for MagAO-X, a relatively short a 5-minute closed-loop period reduces shape creep to $\sim 1$ nm RMS over an hour. Shape drift due to thermal variations in the device temperature may dominate over shape creep and require periodically re-closing the loop on the NCPC DM, although no significant NCPC drift was noticed during MagAO-X on-sky commissioning activities.

\section{Instrument integration and optimization}
\label{sect:instrument}

The DMs were integrated with MagAO-X in early 2019 and aligned using an internal telescope simulator. \rev{The telescope simulator comprises a supercontinuum laser coupled into a single-mode fiber to create a point source that is imaged through a lens to form a collimated beam at a pupil stop etched with the Magellan Clay aperture. A second lens after the pupil creates a f/11 input beam to match the telescope input.} To minimize the static instrumental aberrations that remained after this alignment and to handle system drift over time, we implemented a simple algorithm (the so-called ``eye doctor''\cite{Bailey2014}) to maximize the instrument Strehl ratio. The core of the PSF is measured in a focal plane as a DM iterates over a precomputed modal basis set. As the amplitude of each mode is varied in a grid search pattern, a quadratic polynomial is fit to the measured core to estimate the amplitude that maximizes the Strehl ratio. To compensate for ALPAO creep (particularly immediately after power-on, when creep-induced WFE is most dramatic), modes may be revisited multiple times. To create an orthogonal basis set over each DM's illuminated aperture, the illuminated  DM surface is estimated by measuring the intensity response to a Hadamard modal basis on a defocused PSF and then thresholding on the row-wise RMS of the reconstructed IFs. The RMS response of the $n^\mathrm{th}$ actuator is given by
\begin{align}\label{eqn:illum}
R_n = \mathrm{RMS} \{ (\mathrm{W} \mathrm{H}^{-1})_n \}	 = \frac{1}{N} \sqrt{\frac{1}{M} \mathrm{diag}({ \mathrm{H} \mathrm{W^T} \mathrm{W} \mathrm{H^T})_n}}
\end{align}
where $\mathrm{H}$ is an $N\times N$ Hadamard matrix, we've used the property $\mathrm{H^{-1}} = \frac{1}{N}\mathrm{H^T}$, and $\mathrm{W}$ is the $M\times N$ matrix of measured intensities, computed from difference of the PSF response to positive and negative Hadamard modes $\mathrm{W} = (\mathrm{W}_+ - \mathrm{W}_-)/2$. The modal basis set is then formed by orthogonalizing a set of Zernike polynomials over the illuminated actuators \rev{via the Gram-Schmidt process, as implemented in \texttt{poppy}\cite{poppy},} and interpolating outside the beam footprint from an average of nearest illuminated neighbors.

\begin{figure} 
\begin{center}
\begin{tabular}{c}
\includegraphics{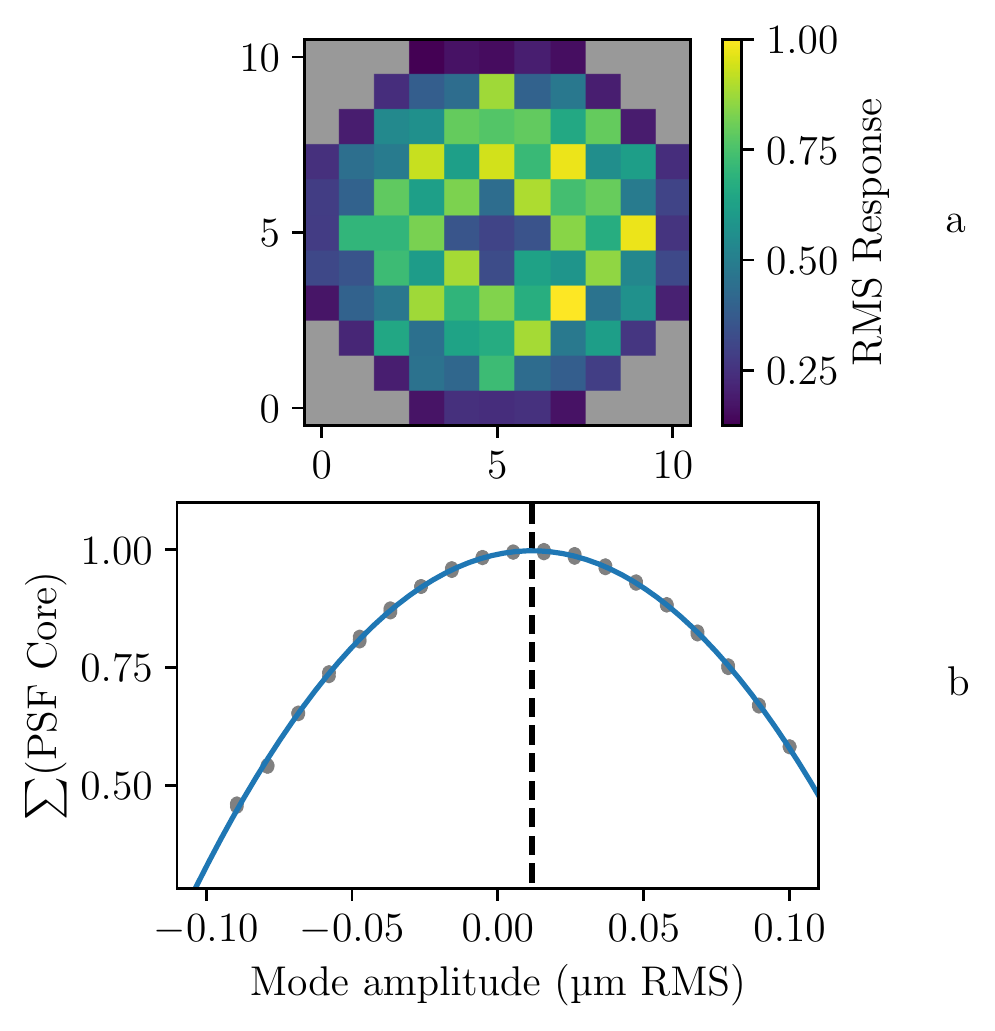}
\end{tabular}
\end{center}
\caption 
{ \label{fig:eye_doctor}
(a) Estimate of illuminated footprint on the woofer DM, calculated following Equation \ref{eqn:illum}. (b) Single-mode example of Zernike grid optimization process. The normalized intensity of the PSF core (grey dots) is measured as the amplitude of each mode is swept over a range of values. A quadratic fit (blue line) gives the value of the mode amplitude that maximizes the intensity in the PSF core (dashed line). \rev{The particular example shown here is oversampled to demonstrate the procedure.}} 
\end{figure}

\rev{When the instrument requires optimization, all DMs are commanded to their flat states, and a single DM is optimized at a time while the others are held in their flat states. The typical approach is to start by optimizing the woofer DM on measurements in a focal plane picked off right before the pyramid tip, followed by a repeat of this process downstream with the NCPC DM and a science camera or the low-order wavefront sensing (LOWFS) camera. After each DM is optimized, the flat command is updated to incorporate the optimized shape. The tweeter flat is not typically optimized via the eye doctor routine to avoid driving low-orders into the flat and reducing the available stroke.} An example curve showing the computed illumination pattern on the woofer and the sum over the measured PSF core as a function of a single mode amplitude is shown in Figure \ref{fig:eye_doctor}.

To further optimize instrument WFE, we implemented the parametric focus-diversity phase retrieval (FDPR) algorithm described in Thurman et al. (2009)\cite{Thurman2009} and demonstrated on the HiCAT bench \cite{Brady2019}. This algorithm attempts to minimize the error between a set of irradiances measured in $K$ defocused planes and a set of modeled PSFs propagated to these planes by the angular spectrum approach. The objective function is given by\cite{Thurman2009}
\begin{align}\label{eqn:obj}
	\Phi = 1 - \frac{1}{K} \sum_{k=1}^K \frac{ \bigg[ \sum_{(x,y)} W_k(x,y) \hat{G}_k(x,y) G_k(x,y) \bigg]^2 } { \bigg[ \sum_{(x,y)} W_k(x,y) G_k(x,y) \bigg]^2 \bigg[ \sum_{(x,y)} W_k(x,y) \hat{G}_k(x,y) \bigg]^2},
\end{align}
 a weighted measure of the correlation between the modeled and measured PSFs, where $G_k(x,y)$ is the measured irradiance in each defocused plane, $\hat{G}_k(x,y)$ is the corresponding modeled irradiance pattern, and $W_k(x,y)$ is a pixel-wise weighting for each plane.
 
The model is parametrized in the axial and longitudinal positions of the PSFs, modal amplitude and phase values, pixel-by-pixel amplitude and phase values, and can enforce a set of smoothness and edge constraints on the pupil solution. Thurman et al. (2009)\cite{Thurman2009} give the analytical form of the gradient terms for each of the parameters. We followed a similar approach to HiCAT\cite{Brady2019}: first optimizing over the longitudinal and axial positions of the PSFs, followed by optimization over the first 45 Zernike modes in phase, and finally multiple optimization steps over pixel-by-pixel amplitude and phase---beginning with a broad Gaussian smoothing kernel in the pupil and reducing the smoothing with each step. We computed a weighting from the measured PSFs as the inverse of the measured irradiance (equivalent to the inverse variance of a Poisson distribution), with a regularization term to compensate for the near-zero background pixels:
\begin{align}
W_k(x,y) = \frac{1} {\mathscr{S}\{G_k(x,y)\} + \epsilon}
\end{align}
where $\mathscr{S}$ is a Gaussian smoothing operator. We minimized Equation \ref{eqn:obj} with a limited memory Broyden–Fletcher–Goldfarb–Shanno bound-constrained (L-BFGS-B) optimizer\cite{2020SciPy-NMeth,LBFGSB1,LBFGSB2} and computed the forward propagation model, objective function, and Jacobian terms on a graphics processing unit (GPU)\cite{cupy_learningsys2017}.

\begin{figure}
\begin{center}
\begin{tabular}{c}
\includegraphics{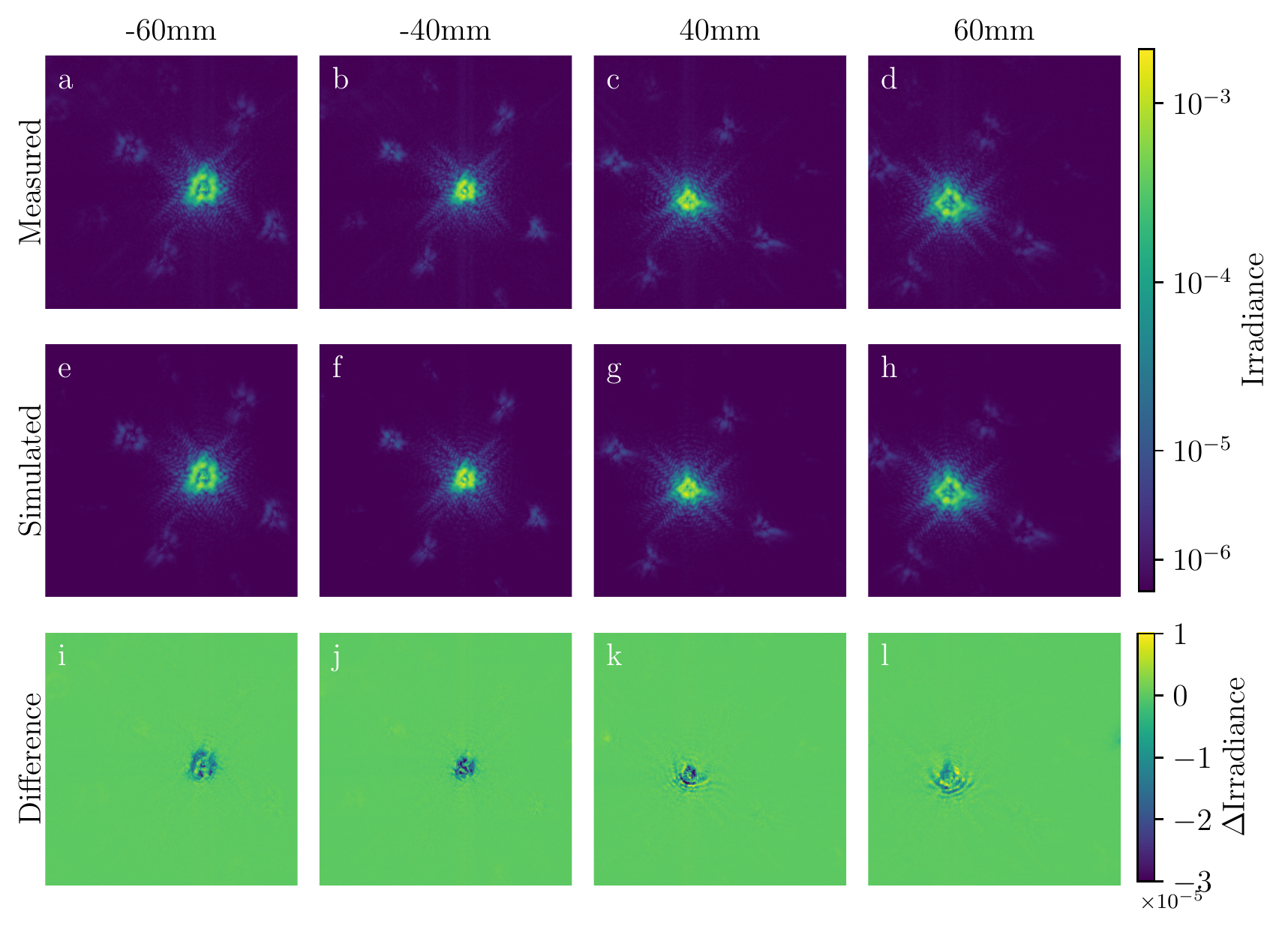}
\end{tabular}
\end{center}
\caption 
{ \label{fig:defocused_psfs}
Top (a,b,c,d): Measured focus-diversity PSFs captured on the science camera over a 120mm longitudinal range. Middle (e,f,g,h): Simulated PSFs produced by the FDPR optimization to these measured data. \rev{Irradiance values are displayed on a logarithmic stretch to reveal PSF structure and normalized such that the total irradiance is equal to unity. Bottom (i,j,k,l): Differences of the measured and simulated PSFs.}} 
\end{figure}

We implemented FDPR on MagAO-X with the internal source and telescope simulator and measured defocused PSFs on a science camera in an f/69 beam in an H$\mathrm{\alpha}$ filter (9 nm bandpass centered at 656 nm) over 120mm of axial motion. Examples of these measurements and the retrieved modeled PSFs are shown in Figure \ref{fig:defocused_psfs}.
 
\begin{figure}
\begin{center}
\begin{tabular}{c}
\includegraphics{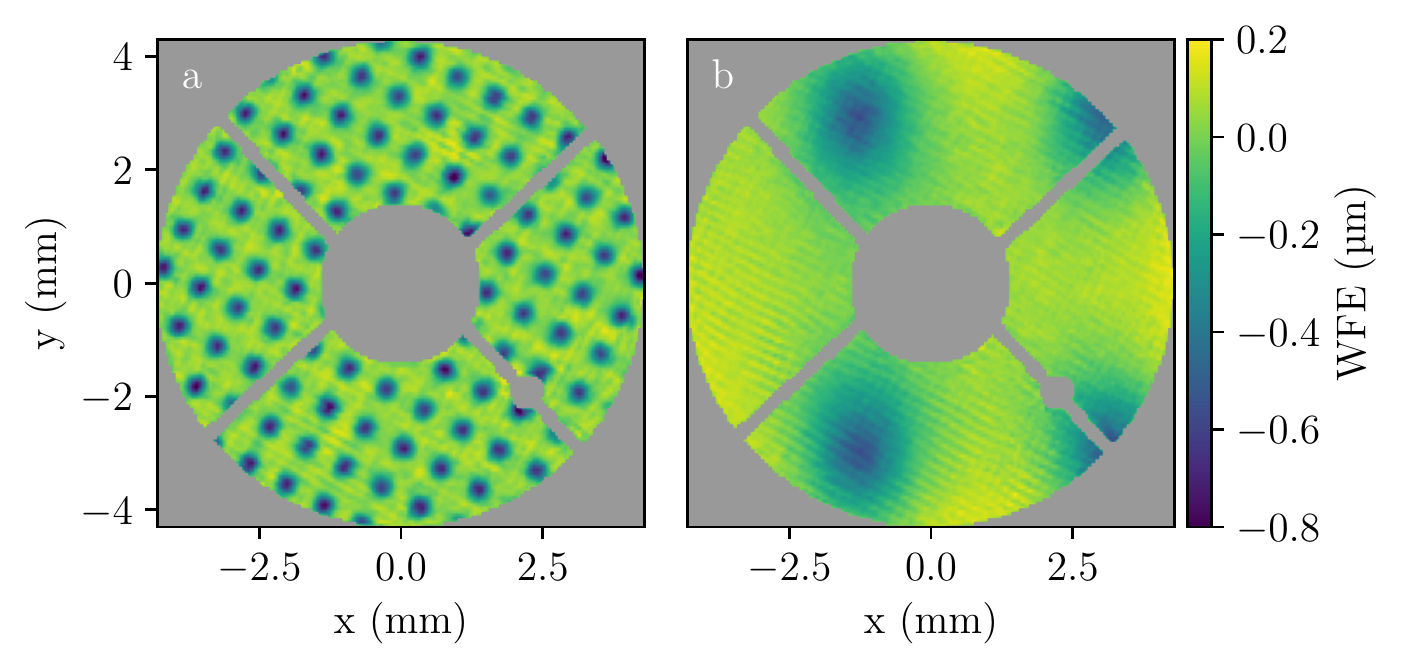}
\end{tabular}
\end{center}
\caption 
{ \label{fig:fdpr_grids}
Phase estimates obtained via FDPR with every 4th actuator poked on (a) the tweeter and (b) the NCPC DM.} 
\end{figure}

We followed the procedure described in Section \ref{sect:char} to build an interaction matrix from pupil phase estimates of grid patterns of poked actuators on the DMs. Examples of the retrieved grid phases on the tweeter and NCPC DMs are shown in Figure \ref{fig:fdpr_grids}. To optimize the instrument aberrations, we closed the loop around the phase retrieved via FDPR, starting with the NCPC DM to drive out the low-order aberrations before switching to the tweeter to drive out the remaining high-orders. The Strehl ratio is estimated in two ways. The first is calculated from the pupil-plane electric field\cite{Hardy1998, Roberts2004}, and the second from focal-plane quantities:
\begin{align}
S_1 = e^{-\sigma^2_{\hat{\phi}}} e^{-\sigma^2_{\hat{\chi}}}\label{eqn:strehl1} \\
S_2 = \frac{I(x_0,y_0)} { \mathscr{S} \{ I_0(x_0,y_0) \} } \label{eqn:strehl2}
\end{align}
where $\sigma^2_{\hat{\phi}}$ is the variance of the estimated pupil-plane phase, $\sigma^2_{\hat{\chi}}$ is the variance of the estimated pupil-plane log-amplitude, $I(x_0,y_0)$ is the peak value of the measured PSF, and $I_0(x_0,y_0)$ is the peak of the aberration-free polychromatic PSF simulated with the same subpixel alignment over the H$\alpha$ filter bandpass. $\mathscr{S}$ is a Gaussian smoothing operator to account for the effects of detector charge diffusion and tip/tilt vibrations happening faster than the integration time. We derive the expression for the Strehl ratio $S_1$ in Appendix \ref{sect:appendix}.

\begin{figure}
\begin{center}
\begin{tabular}{c}
\includegraphics{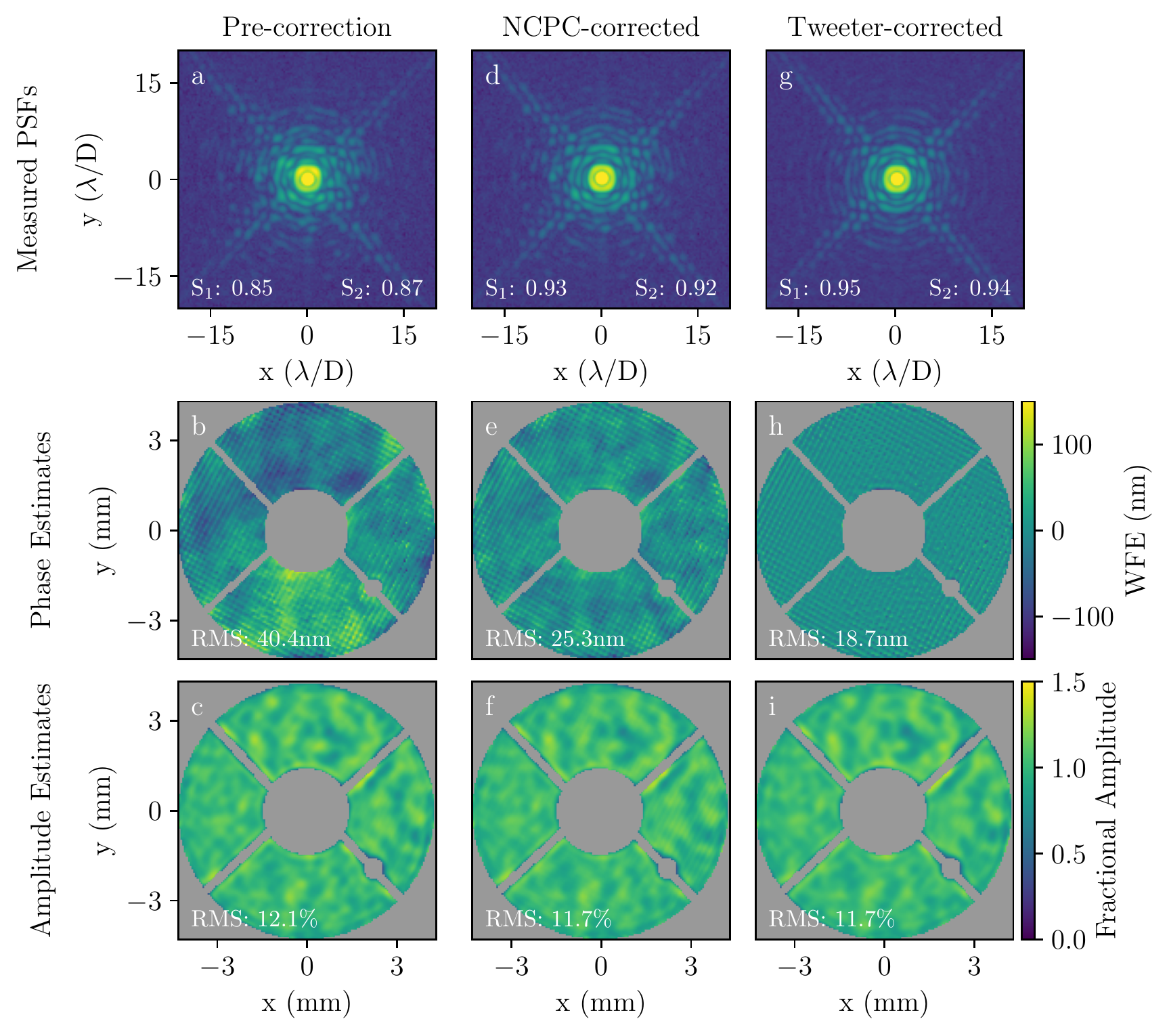}
\end{tabular}
\end{center}
\caption 
{ \label{fig:fdpr_psfs_wfes}
Left column: (a) Measured PSF at H$\mathrm{\alpha}$, (b) estimated phase, and (c) estimated amplitude prior to correction with FDPR. Middle column: (d) Measured PSF at H$\mathrm{\alpha}$, (e) estimated phase, and (f) estimated amplitude after closed-loop correction with the NCPC DM. Right Column: (g) Measured PSF at H$\mathrm{\alpha}$, (h) estimated phase, and (i) estimated amplitude after closed-loop correction with both the NCPC and tweeter DMs. Strehl ratios are reported for each measured PSF, following Equations \ref{eqn:strehl1} and \ref{eqn:strehl2}. RMS WFE values are reported for each phase estimate, and percentage amplitude RMS is reported for each amplitude estimate.}
\end{figure}

\begin{figure}
\begin{center}
\begin{tabular}{c}
\includegraphics{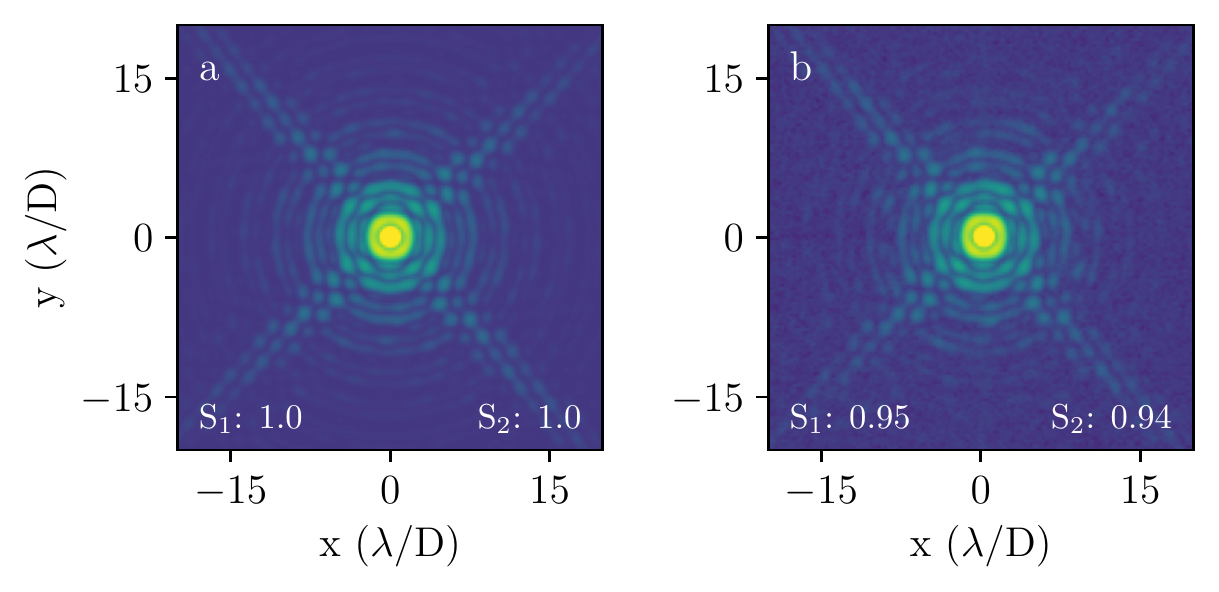}
\end{tabular}
\end{center}
\caption 
{ \label{fig:psf_sim_0WFE} (a) Simulated 0-WFE PSF in the H$\alpha$ bandpass, convolved with a 0.6$\sigma$ pixel Gaussian smoothing kernel to account for the effect of detector charge diffusion. (b) Measured PSF at H$\alpha$ after correction with the NCPC and tweeter DM. Panel (b) is repeated from Figure \ref{fig:fdpr_psfs_wfes} (g).}
\end{figure}

 These results are summarized in Figures \ref{fig:fdpr_psfs_wfes} and \ref{fig:psf_sim_0WFE}. The reported WFE residuals are the estimates obtained from FDPR, and the Strehl ratios are computed following Equations~\ref{eqn:strehl1} and \ref{eqn:strehl2}. \rev{RMS amplitude is computed from the FDPR-derived pupil-plane amplitude estimates, normalized such that the mean amplitude within the pupil is equal to unity to yield a dimensionless measure of the fractional variation of the amplitude across the pupil.} In this case, the starting phase aberration was estimated at 40.4 nm RMS wavefront with an estimated Strehl ratio of 0.85-0.87. The NCPC correction drove these values to 25.3 nm RMS wavefront and a Strehl ratio of 0.92-0.93. The final tweeter correction resulted in an estimated 18.7 nm RMS wavefront error with a Strehl ratio of 0.94-0.95. The variation of amplitude over the pupil is estimated to be 11-12\% RMS, which decreases the Strehl ratio by 1-2 percentage points compared to an estimate that accounts for phase only. In Lumbres et al. (in prep)\cite{lumbresprep}, an end-to-end Fresnel model of the instrument is compared with instrument performance and yields an estimate of the residual WFE and Strehl ratio consistent with the values reported here.

\section{Conclusions}
\label{sect:conc}

We have described our procedure for and the results of characterizing the DMs for the ExAO instrument MagAO-X. The 2K BMC DM serves as the instrument tweeter, with a best flat of $8.1 \si{\nano\meter}$ RMS surface or $0.5$ nm RMS surface in-band over the coronagraphic pupil. The two ALPAO DMs, one of which serves as the woofer and the other as a non-common-path correction downstream of the main AO loop, can be flattened to 2.6 and 4.9 nm RMS surface, or 0.9 and 1.3 nm RMS surface within their respective control bandpasses and the vAPP pupil. By offloading low-order sag from tweeter to woofer, $>99\%$ of tweeter actuators exceed $0.85\si{\micro\meter}$ of available inter-actuator surface stroke. \rev{By comparison, in the MEMS characterization work undertaken in support of GPI development, a 1K BMC device was flattened to 6.4nm RMS surface and 0.27nm RMS surface in-band across a 27x27 actuator region on the DM\cite{Evans2006}. Additionally, a 4K BMC device was flattened to 5.15 nm RMS surface and 0.26 nm RMS surface in-band across a 30x30 actuator region\cite{norton2009}.}

\rev{Our observation of creep in the ALPAO DM97 devices is consistent with results reported on NAOMI\cite{LeBouquin2018} with a DM241, the DM97s in the Raven instrument\cite{Lamb2015}, and in the extensive characterization and compensation of the creep effect performed by Bitenc et al.\ (2014) \cite{Bitenc2014} and Bitenc (2017)\cite{Bitenc2017} with DM241 and DM277 devices.}

\rev{After integration of the DMs with MagAO-X}, laboratory efforts showed that the Strehl ratio at the science cameras can be optimized to an estimated 94-95$\%$ and the instrument wavefront error reduced to 18.7 nm RMS, via brute-force optimization of low-order Zernikes followed by focus-diversity phase retrieval on the science cameras. The speed of the current implementation of FDPR is limited by camera stage movement and integration time at each position. Future efforts in this area will investigate using MagAO-X's two science cameras and dedicated LOWFS camera to enable simultaneous focus-diverse measurements, an approach that could allow FDPR to be used for NCP WFE correction on-sky.

\appendix
\section{Estimating Strehl ratio in the presence of scintillation}
\label{sect:appendix}

\rev{We derive an expression for the Strehl ratio that incorporates amplitude aberrations, stated in this manuscript as $S_1$ in Equation \ref{eqn:strehl1}. This result is well known and stated by others\cite{Roberts2004}, but here we provide a derivation in terms of the probabilistic method of Ross (2009)\cite{ross2009}, which we believe has not appeared in the literature.}

Given a pupil-plane electric field of the form $E(\bm{r}) =  A(\bm{r}) e^{i\phi(\bm{r})}$, the irradiance in the focal plane can be written as $I(\bm{r_f}) = | \mathscr{F}\{ A(\bm{r})e^{i \phi(\bm{r})} \} |^2$, where $A(\bm{r})$ is the amplitude over the aperture and $\phi(\bm{r})$ is the phase. The Strehl ratio is defined as
\begin{align}
S = \frac{I(\bm{0})} {I_0(\bm{0})},
\end{align}
the ratio of the on-axis irradiance of the aberrated field to the on-axis irradiance of the aberration-free field. This expression can be expanded as
\begin{align}
S = \frac{\big| \mathscr{F}\{ E(\bm{r}) \} \big|^2} {\big| \mathscr{F}\{ E_\mathrm{0}(\bm{r}) \} \big|^2} \Bigg|_{\bm{r_f}=\bm{0}} = \frac{\bigg| \displaystyle\int\limits_{\mathrm{aperture}} E(\bm{r}) \dif^2\bm{r} \bigg|^2} {\bigg| \displaystyle\int\limits_{\mathrm{aperture}} E_0(\bm{r}) \dif^2\bm{r} \bigg|^2} = \left|\frac{\displaystyle\int\limits_{\mathrm{aperture}} E(\bm{r}) \dif^2\bm{r} } {\bar{A} \displaystyle\int\limits_{\mathrm{aperture}} \dif^2\bm{r}} \right|^2,
\end{align}
where the second equality is justified by the central ordinate theorem for Fourier transforms, and the aberration-free field $E_0(\bm{r})$ is taken to have a constant amplitude $\bar{A}$ over the aperture and phase $\phi(\bm{r})=0$. In a slight modification to the formulation in Ross (2009)\cite{ross2009}, we rewrite this in terms of the expectation value of the aberrated field to get
\begin{align}\label{eqn:strehl_exp}
	S = \bigg| \frac{1} {\bar{A}} <A(\bm{r})e^{i \phi(\bm{r})}> \bigg|^2.
\end{align}

If the probability distribution function (PDF) of the pupil-plane field is known, an expression for the Strehl ratio can be found directly. Setting \nolinebreak{$A(\bm{r})=\bar{A}$} in the absence of amplitude aberrations, this reduces to \nolinebreak{$S = | <e^{i\phi(\bm{r})}> |^2$}. If the phase follows a Gaussian PDF, Ross (2009)\cite{ross2009} showed that this yields the familiar exponential estimate of the Strehl ratio in terms of the variance of the phase, \nolinebreak{$S=e^{-\sigma^2_\phi}$}, known as the extended Mar{\'{e}}chal approximation\cite{Hardy1998}.

In the presence of amplitude aberrations but perfect phase, Equation \ref{eqn:strehl_exp} becomes
\begin{align}\label{eqn:strehl_amp_nonred}
S = \bigg| \frac{1} {\bar{A}} <A(\bm{r})> \bigg|^2.
\end{align}

Scintillation from atmospheric turbulence has been shown to be well-represented by a log-normal PDF\cite{Fried67,Andrews2005,JuradoNavas2011} of the form
\begin{align}\label{eqn:lognormal}
\mathrm{PDF}_I(I) =  \frac{1}{2 I \sqrt{2\pi \sigma_\chi^2}} \exp\Bigg[{-\frac{[\ln({I/\bar{I}}) + 2\sigma_\chi^2]^2}{8\sigma_\chi^2}}\Bigg],
\end{align}
where $I$ is the irradiance, $\bar{I}$ is the irradiance in the absence of turbulence, and $\sigma_\chi^2$ is the variance of the log-amplitude. With a log-normal PDF for the irradiance, the expression for the Strehl ratio in Equation \ref{eqn:strehl_amp_nonred} becomes
\begin{align}
\begin{aligned}
	S =& \Bigg| \frac{1}{\bar{A}} \int_{-\infty}^{\infty} \frac{1}{2 \sqrt{2 I \pi \sigma_\chi^2}} \exp\Bigg[{-\frac{[\ln{(I/\bar{I})} + 2\sigma_\chi^2]^2}{8\sigma_\chi^2}}\Bigg] \dif I \Bigg|^2 = \Big| e^{-\sigma_\chi^2/2} \Big|^2 \\
	=& e^{-\sigma_\chi^2},
\end{aligned}
\end{align}
where we've used $A=\sqrt{I}$ and $\bar{A} = \sqrt{\bar{I}}$.

The more general case of simultaneous phase and amplitude aberrations requires solving an equation of the form
\begin{align}
S = \Bigg| \frac{1}{\sqrt{\bar{I}}} \int_{-\infty}^{\infty} \int_{-\infty}^{\infty} \sqrt{I} e^{i \phi } \mathrm{PDF_{I, \phi}}(I, \phi) \dif I \dif \phi \Bigg|^2,
\end{align}
where $\mathrm{PDF_{I, \phi}}(I, \phi)$ is the joint probability distribution for the irradiance and phase. If the irradiance and phase are assumed to be statistically independent, this simplifies to
\begin{align}
	S = \bigg| \frac{1} {\bar{A}} <A(\bm{r})>  <e^{i\phi(\bm{r})}> \bigg|^2 = e^{-\sigma_\phi^2} e^{-\sigma_\chi^2},
\end{align}
the form given in Equation \ref{eqn:strehl1}.

\subsection*{Disclosures}

The authors have no conflicts of interest to declare.

\acknowledgments 
We are grateful for the support of NSF MRI Award \#1625441 \textit{Development of a Visible Wavelength Extreme Adaptive Optics Coronagraphic Imager for the 6.5 meter Magellan Telescope}.


\bibliography{report}   
\bibliographystyle{spiejour}   


\vspace{2ex}\noindent\textbf{Kyle Van Gorkom} is a PhD student at the Wyant College of Optical Sciences at the University of Arizona and an optical engineer at NASA Goddard Space Flight Center. His research focuses on adaptive optics for the direct imaging of exoplanets and metrology for future space missions.

\vspace{1ex}
\noindent Biographies and photographs of the other authors are not available.

\listoffigures
\listoftables

\end{spacing}
\end{document}